\documentclass[11pt]{article}
\usepackage{moriond,epsfig,graphicx,amssymb,amsmath}

\def\MyPhoto{\includegraphics[width=35mm]{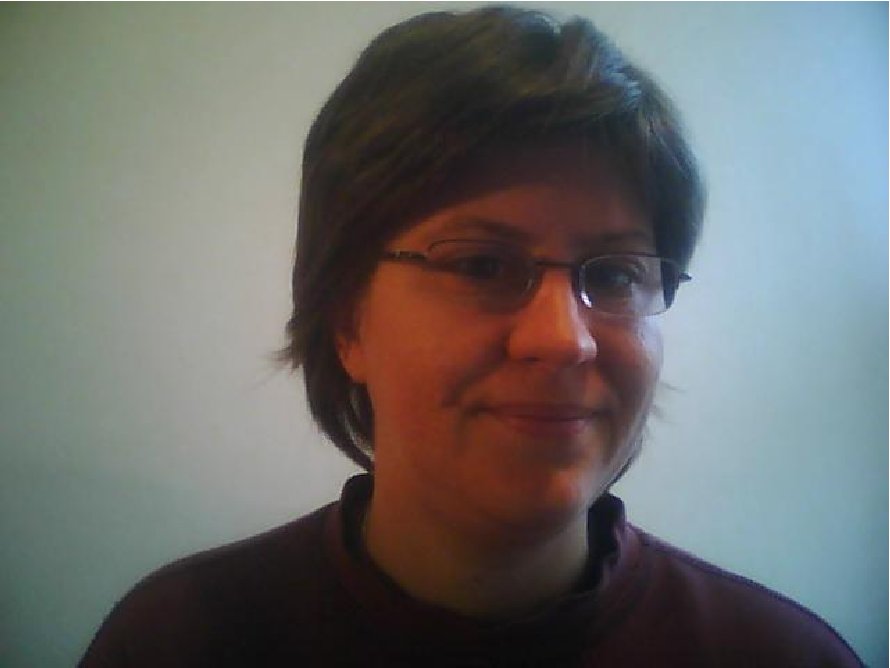}}

\def\be{\begin{equation}}
\def\ee{\end{equation}}
\def\bea{\begin{eqnarray}}
\def\eea{\end{eqnarray}}

\begin{document}
\DeclareGraphicsExtensions{.pdf,.eps,.epsi,.jpg}
\vspace*{4cm}
\title{STATUS REPORT ON DOUBLE CHOOZ}

\author{ A. PORTA for the Double Chooz collaboration}

\address{CEA-Saclay, DSM/IRFU/SPhN Bat. 703\\
91191 Gif-Sur-Yvette, France}

\maketitle
\abstracts{Double Chooz main target is to measure $\theta_{13}$ oscillation
parameter by comparing reactor neutrino fluxes in two identical detectors 
located respectively at 400 m and 1 km away from the 2 Chooz reactor cores. 
The far detector is now under construction, while we have just completed the 
design phase of the near one. In this report I will discuss the detector 
principle, sensitivity and its present construction status.}

\section{Introduction}

The $\theta_{13}$ mixing angle is one of the still completely unknown 
parameters in the frame of neutrino oscillation theory~\cite{teoria}.  
The measurement of its value via the detection of reactor 
$\bar{\nu}_e$ flux is the target of the Double Chooz project~\cite{DC}. 
Present limit on this parameter has been set by the Chooz 
experiment~\cite{chooz1} at $sin^22\theta_{13} < 0.15$ with 90$\%$ C.L. for $\Delta m^2_{31}=2.5 \times 10^{-3} eV^2$.
The Chooz measured fraction of surviving $\bar{\nu}_e$ at 1.05 km from the 
reactor cores is $R=1.01 \pm 2.8\% (stat.) \pm 2.7\% (syst.)$.\\
The main source of systematic error in the Chooz experiment 
was due to the uncertanties on the original neutrino flux emitted by the 
reactors. To reduce this error the Double Chooz experiment will use 2 
identical detectors: the far one (d $\simeq$ 1.05 km), placed in the same 
laboratory of the first Chooz experiment, to measure the oscillation effect on 
neutrino flux and a near one (d $\simeq$ 400 m) to measure the absolute 
neutrino flux emitted by the reactors.
The detectors will also have an improved geometry to maximize the target 
volume and the background suppression.

\section{The detectors}
\begin{figure}[h]
\begin{center}
\includegraphics[height=2.5in]{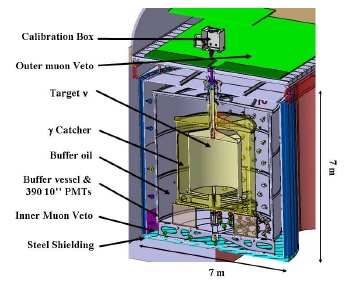}
\caption{Double Chooz detector design.
\label{fig:DC}}
\end{center}
\end{figure}

Reactor $\bar{\nu}_e$'s are detected by using a Gd-doped liquid scintillator 
target. The detection reaction is the inverse beta decay interaction: 
$\bar{\nu}_e + p \to e^+ + n$. 
The $e^+$ detection produces a prompt signal of energy 
$E_{prompt}=E_{\bar{\nu}_e}-(M_n-M_p)+m_e$. This first pulse is followed 
by a delayed signal induced by the radiative capture of the neutron on Gd 
with the emission of a gamma cascade of total energy $\sim$ 8 MeV. 
The characteristic neutron capture time is $\sim$30 $\mu$s.\\ 
The Double Chooz detector design has been made to maximize $\bar{\nu}_e$ 
detection efficiency and to reduce the background which can imitate 
this characteristic double signal reaction.\\
The detector consists of 4 concentric liquid volumes (figure \ref{fig:DC}).
Starting from the center we have the target made by 10 $m^3$ of Gd-doped liquid 
scintillator contained in an acrylic vessel. The target is surrounded by a 55 
cm-thick $\gamma$-catcher of non-loaded liquid scintillator also contained in 
an acrylic vessel. This geometry maximizes the detection efficiency of the 
$\gamma$'s emitted by $e^+$ annihilation and neutron capture on Gd.
This inner part is observed by 390 photomultipliers immersed in a 105 cm thick
non-scintillating mineral oil buffer designed to shield the detector core from 
the natural radioactivity of the phototubes.  
This buffer is contained in a steel tank surrounded by a 50 cm-thick liquid scintillator 
muon veto (inner muon veto) and by 15 cm of steel to shield rock radioactivity. 
An extra plastic scintillator muon veto (outer muon veto) is placed on the top 
of the detector to tag the muons crossing the rock around the detector.

\section{Sensitivity to $\theta_{13}$}

\begin{figure}[h]
\begin{center}
\includegraphics[height=2.5in]{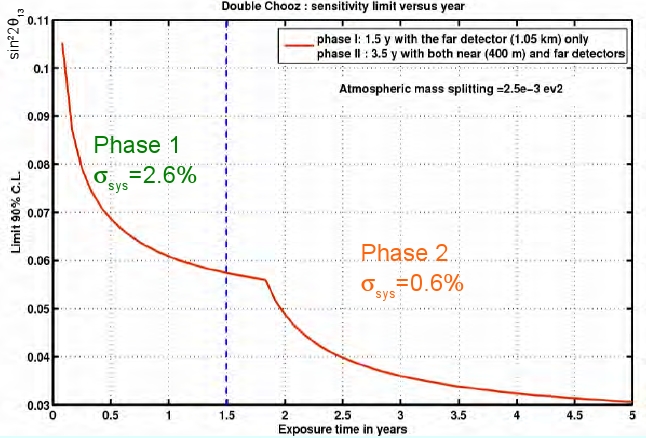}
\caption{Double Chooz expected sensitivity limit (90$\%$ C.L.) to $sin^22\theta_{13}$ as a function of time. Phase 1 (far detector only) will last 1.5 years, phase 2 (both detectors) will last 3 years.
\label{fig:sens}}
\end{center}
\end{figure}

The Double Chooz project foresees 2 phases: a first one starting at the 
beginning of 2010 and lasting $\sim$1.5 years with only the far detector 
running and a second one, lasting $\sim$3 years, with the 2 detectors 
taking data together.\\
The absolute (phase 1) and relative (phase 2) systematic uncertainties~\cite{errori} on the antineutrino flux measured by Double Chooz are summarized in 
table \ref{tab:uncert}.
\begin{table}[t]
\caption{Absolute and relative errors on the antineutrino flux measurement performed by the Double Chooz experiment.\label{tab:uncert}}
\vspace{0.4cm}
\begin{center}
\begin{tabular}{|c|c|c|}
\hline
Error description & Absolute & Relative\\
\hline
Production cross section & 1.9$\%$ & /\\
\hline
Reactor power & 2.0$\%$ & /\\
\hline
Energy per fission & 0.6$\%$ & /\\
\hline
Solid angle & / & 0.06$\%$\\
\hline
Detection cross section & 0.1$\%$ & /\\
\hline
Target mass & 0.2$\%$ & 0.2$\%$\\
\hline
Number of protons & 0.5$\%$ & 0.1$\%$\\
\hline
Particle identification & 0.4$\%$ & 0.4$\%$\\
\hline
\end{tabular}
\end{center}
\end{table}
The expected Double Chooz sensitivity limit (90$\%$ C.L.) to 
$sin^22\theta_{13}$ during the 2 experimental phases (figure \ref{fig:sens})
has been calculated by taking into account the errors of table 
\ref{tab:uncert}.\\ 
Double Chooz will reach the Chooz limit after about one month of data taking 
with only the far detector running. At the end of the first phase 
we expect a sensitivity  of $sin^22\theta_{13} > 0.06$ with 90$\%$ C.L. for 
$\Delta m^2_{31}=2.5 \times 10^{-3} eV^2$. At the end of the second phase, 
we will reach a sensitivity of $sin^22\theta_{13} > 0.03$ with 90$\%$ C.L. for 
$\Delta m^2_{31}=2.5 \times 10^{-3} eV^2$.

\section{Construction status and prospects}
\begin{figure}[]
\begin{center}
\includegraphics[height=2.5in]{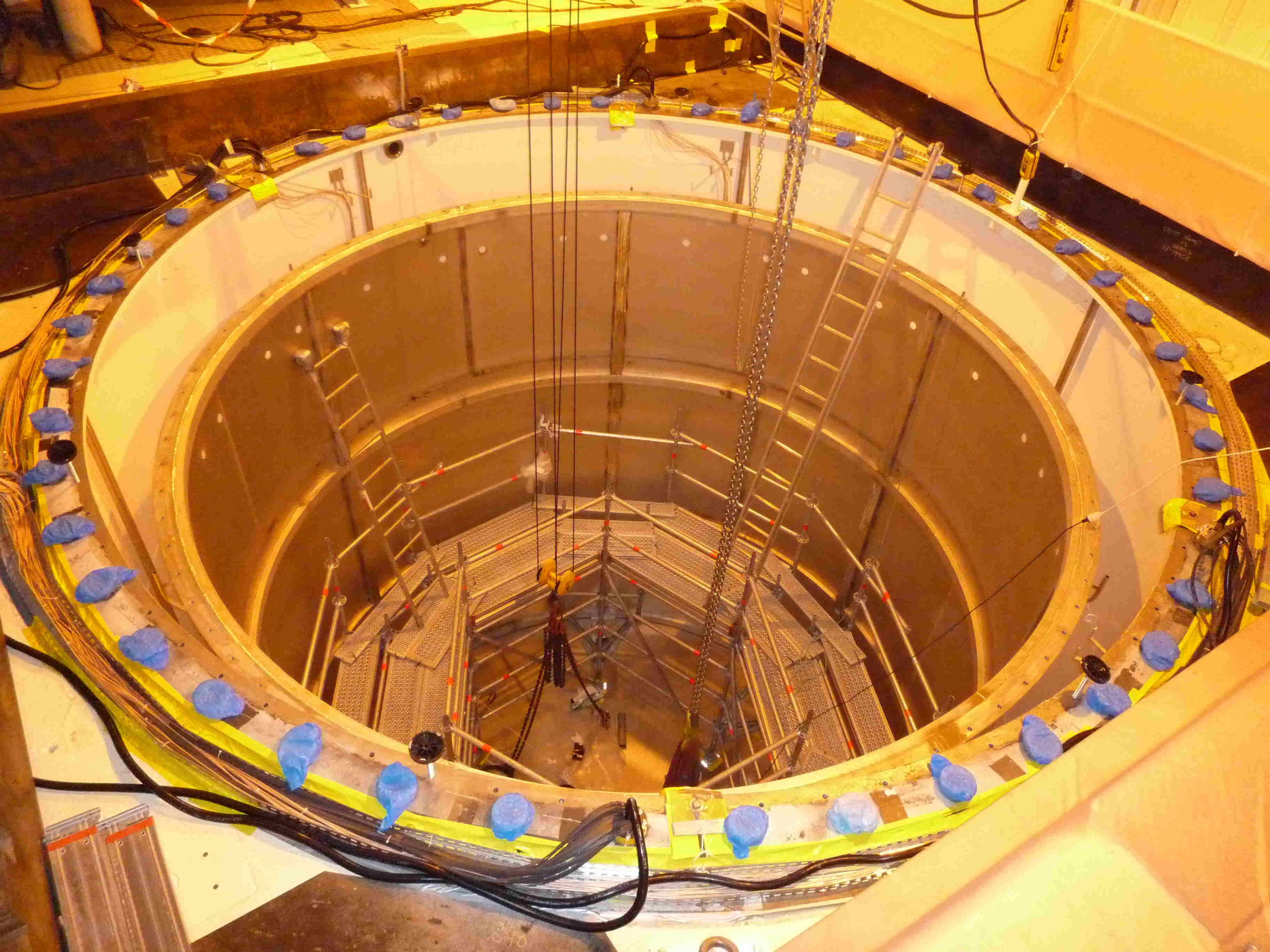}
\caption{Buffer tank installation.
\label{fig:buffer}}
\end{center}
\end{figure}

\begin{figure}[h]
\begin{center}
\includegraphics[height=2.5in]{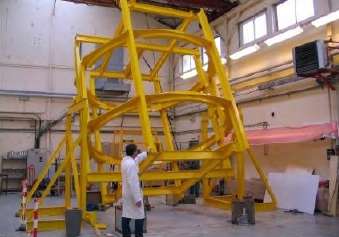}
\caption{Mechanical structure for the transportation and installation of the  $\gamma$-catcher vessel in the detector.
\label{fig:vessel}}
\end{center}
\end{figure}

The far detector construction started in May 2008 with the integration of the 
external steel shielding. The integration of the inner veto tank has been 
accomplished in autumn 2008.
After a preliminary phase for painting and cleaning the laboratory and veto tank, last February we successfully
completed the veto photomultiplier installation. By the beginning of May 2009 we plan to finish the buffer tank
integration and cleaning (figure \ref{fig:buffer}) and to start the installation of the 390 buffer phototubes.\\
During the summer we will proceed with the $\gamma$-catcher and target acrylic 
vessel integration by means of 2 special structures expressly designed to 
build, transport and install them in the laboratory pit 
(figure \ref{fig:vessel}).\\
Meanwhile a liquid storage area is under construction near the Chooz 
laboratory. The 4 liquids will be delivered to this area in September. In 
autumn we will install the DAQ electronics and we will perform the detector 
filling. During this phase we will measure with high 
precision the target liquid weight. This will allow us to maximally reduce 
the uncertainty in the calculation of the number of target free protons.\\ 
The detection of the first neutrino event is foreseen for the end of 2009. 
The far detector installation will be completed at the beginning of 2010 with 
the integration of the outer veto and of the glove box for calibration source 
manipulation.\\
The design of the near laboratory has been already completed and the end of 
its construction is foreseen for middle 2011 to have the detector ready by the 
end of 2011.

\section{Conclusions}

Double Chooz is the first new generation reactor neutrino detector to use 
the 2 identical detector technique to measure the lack in neutrino flux 
induced by the $\theta_{13}$ mixing angle.
The far detector, now under construction, will start to take data at 
the beginning of 2010. After 1.5 years of data taking with only the far 
detector we will reach a sensitivity of $sin^22\theta_{13} > 0.06$ 
(90$\%$ C.L).
At the end of 2011 the near detector will be finished and, with the 2 detectors running together, we expect to reach a sensitivity of $sin^22\theta_{13} > 0.03$ (90$\%$ C.L) within 3 years.\\
The Double Chooz results will integrate and complete future accelerator 
measurements allowing to better constrain the 2 still unknown oscillation 
parameters: the $\theta_{13}$ mixing angle and the CP violation phase.

\end{document}